# Observation of Dimension-Crossover of a Tunable 1D Dirac Fermion in Topological Semimetal $NbSi_xTe_2$

Jing Zhang[1,2], Yangyang Lv[3], Xiaolong Feng[4], Aiji Liang[1,5], Wei Xia[1,5], Sung-Kwan Mo[6], Cheng Chen[1,6,7], Jiamin Xue[1], Shengyuan A. Yang[4], Lexian Yang[8], Yanfeng Guo[1,5], Yanbin Chen[3], Yulin Chen[1,5,7,‡✉], Zhongkai Liu[1,5,†✉]

[1]*School of Physical Science and Technology, ShanghaiTech University, Shanghai 201210, China*

[2]*University of Chinese Academy of Sciences, Beijing 100049, China*

[3]*National Laboratory of Solid-State Microstructures, School of Physics and Collaborative Innovation Centre of Advanced Microstructures, Nanjing University, Nanjing 210093, China*

[4]*Research Laboratory for Quantum Materials, Singapore University of Technology and Design, Singapore 487372, Singapore*

[5]*ShanghaiTech Laboratory for Topological Physics, Shanghai 201210, China*

[6]*Advanced Light Source, Lawrence Berkeley National Laboratory, Berkeley, CA 94720, USA*

[7]*Department of Physics, University of Oxford, Oxford, OX1 3PU, UK*

[8]*State Key Laboratory of Low Dimensional Quantum Physics, Department of Physics, Tsinghua University, Beijing 100084, China*

‡yulin.chen@physics.ox.ac.uk

†liuzhk@shanghaitech.edu.cn

**Condensed matter systems in low dimensions exhibit emergent physics that does not exist in three dimensions. When electrons are confined to one dimension (1D), some significant electronic states appear, such as charge density wave, spin-charge separations and Su-Schrieffer-Heeger (SSH) topological state. However, a clear understanding of how the 1D electronic properties connects with topology is currently lacking. Here we systematically investigated the characteristic 1D Dirac fermion electronic structure originated from the metallic $NbTe_2$ chains on the surface of the composition-tunable layered compound $NbSi_xTe_2$ ($x$ = 0.40 and 0.43) using angle-resolved photoemission spectroscopy. We found the Dirac fermion forms a Dirac nodal line structure protected by the combined $\widetilde{M}_y$ and time-reversal symmetry $T$ and proves the $NbSi_xTe_2$ system as a topological semimetal, in consistent with the ab-initio calculations. As $x$ decreases, the interaction between adjacent $NbTe_2$ chains increases and Dirac fermion goes through a dimension-crossover from 1D to 2D, as evidenced by the variation of its Fermi surface and Fermi velocity across the Brillouin zone in consistence with a Dirac SSH model. Our findings demonstrate a tunable 1D Dirac electron system, which offers a versatile platform for the exploration of intriguing 1D physics and device applications.**

## INTRODUCTION

Low dimensional electronic states bear fascinating emergent physics and have attracted great research interests in the past decade. For example, some of the two-dimensional (2D) electronic states under intensive investigation include electronic structure of van der Waals (vdW) materials in the atomic limit[1-7], the quantum well states confined on the surface/interfaces of semiconductors[1,8,9] and topological surface states of strong topological insulators[10-13]. They host profound physics including Ising superconductivity[14-16], charge density wave[17-20], spin-momentum locking[21,22] and magnetism[23-25]. As electrons are confined into a one-dimensional (1D) wire, fundamentally important phenomena may emerge, such as Peierls phase transition (Perierls instability)[26,27], Su-Schrieffer-Heeger (SSH) topological state[28] and solitonic excitation[29-31], topological edge[32-34]/hinge[35-37] states, as well as Tomonaga-Luttinger liquid behavior[38-40] due to electron correlation. These meaningful properties, together with the topological protection due to the bulk-surface correspondence in topologically nontrivial quantum materials, making low dimensional electronic states candidates for next generation electronic devices.

Despite these great interests, the experimental realization of a tunable 1D electronic states appears to be rare. Physical systems with 1D electron systems include the self-assembly growth of metal wires on semiconductor surface (e.g., In[41], Ge[42,43], Si[44], GaN[45] wires on Si(111), Au wires on Ge(110)[46] and Ge(001)[47]), $MoO_x$[48,49], $Li_{0.9}Mo_6O_{17}$[50], $Rb_{0.3}MoO_3$[51], $TaSe_3$[52], $X$$Te_3$ ($X$ = Nb, V, Ti)[53], $NbSe_3$[54], etc. However, a few of them can be served as a stoichiometric compounds system for research on tunable 1D electronic states[55-57]. For the ease of investigation, manipulation and device fabrication, it would be desirable to search for stoichiometric compounds hosting tunable 1D electronic states.

Recently, the family of composition-tunable compounds $NbSi_xTe_2$ ($x$ = 0.33 to 0.5) or $Nb_{2n+1}Si_nTe_{4n+2}$ ($n$ = 1, 2, ..., ∞) offers an ideal platform for studying 1D electrons. Previous studies have revealed 1D metallic $NbTe_2$ chain structure[58,59], 1D unidirectional massless Dirac fermions ($x$ = 0.45)[60] and nodal lines structure protected by the nonsymmorphic glide mirror symmetry ($x$ = 0.33) [61,62]. By changing $n$ or $x$ values, the spacing between neighboring chains could be altered, which results in the tunable electronic states[63]. The gapless nature,

fast electron velocity (~$10^5$ m·$s^{-1}$), great tunability as well as the thermal and air stability will facilitate its future applications in 1D electronics. However, the 1D nature of the Dirac fermion, its relationship with topology, as well as its evolution with different $x$ values, has not been systematically investigated and carefully explained.

In this work, by using angle-resolved photoemission spectroscopy (ARPES), we systematically studied the electronic structure of $NbSi_xTe_2$ with varying $x$ values. Our result proves the 1D nature of the characteristic Dirac fermion, which forms a nodal line structure along the Brillouin zone (BZ) boundary protected by the nonsymmorphic symmetry, showing a good agreement with the ab-initio calculations. Interestingly, in samples with smaller $x$, the 1D Dirac fermion experiences a dimension-crossover from 1D to 2D as evidenced by the enhanced variation in the Fermi surface and the Dirac fermion velocity across the nodal line. Such dimension-crossover is due to the stronger interactions between the neighboring metallic $NbTe_2$ chains, which could be nicely captured by our proposed Dirac SSH model that incorporates the important nonsymmorphic symmetry constraint into the conventional SSH model. Our observation demonstrates a tunable 1D Dirac fermion which could be utilized in device applications such as fast unidirectional conduction wires.

## RESULTS

### Crystal structure and characterizations of $NbSi_xTe_2$.

The $NbSi_xTe_2$ are vdW crystals with layered structures. Each layer is constituted of three kinds of chain-like structures as building blocks (Fig. 1**a-b**): type I and II are semiconducting $Nb_2SiTe_4$ chains and type III are metallic $NbTe_2$ chains. With the inclusion of the $NbTe_2$ chains, $NbSi_xTe_2$ or $Nb_{2n+1}Si_nTe_{4n+2}$ becomes metallic. With different $n$ or $x$ values, the lattice

structure also changes from the orthorhombic lattice (space group *Pnma*, No. **62**) with $x$ = 0.33 ($n$ = 1) to the monoclinic lattice (space group $P12_1/c1$, No. **14**) with $x$ = 0.50 ($n = \infty$), but the non-symmorphic glide mirror symmetry $\widetilde{M_y}$ perpendicular to the $b$-axis is preserved. The BZ of $x$ = 0.33 ($n$ = 1) is shown in Fig. 1**c**. And STM topography image of $NbSi_xTe_2$ clearly indicates the existence of the metallic chains, which are shown in form of bright lines in Fig. 1**d** (data from Ref. **58**).

The representative electronic bandstructure of $x$ = 0.40 ($n$ = 2) sample could be found in Fig. 1**e-g**. Dispersions across the Fermi level are observed, suggesting its metallic nature. We note a linear dispersion could be found along the $\bar{\Gamma} - \bar{X} - \bar{\Gamma}$ direction (Fig. 1**f**) and the Dirac point forms a flat band along the $\bar{X} - \bar{S}$ direction (Fig. 1**g**), indicating a Dirac nodal line structure protected by the combined $\widetilde{M_y}$ and time-reversal symmetry $T$. Our measurement shows nice agreement with the ab-initio calculations (Fig. 1**h**, also see Ref. **63**) (The spin-orbit coupling effect in this system is fairly weak and only leads to very small band splitting thus could be neglected[61,63], see discussion in Supplementary Figure 1).

**Band structure of the 1D Dirac fermion of $x$ = 0.43 ($n$ = 3) sample.**

We proved the 1D Dirac fermions nature by measuring its dispersion and evolution along all momentum directions (see the schematic of 1D Dirac fermion in Fig. 2**b-c**). The Dirac cone shape is pronounced along the $\bar{\Gamma} - \bar{X} - \bar{\Gamma}$ ($k_x$) direction, as could be observed in the second derivative of the spectrum of $x$ = 0.43 ($n$ = 3) sample (Fig. 2**a**). The estimated Fermi velocity is 1.07 eV·Å (~$1.63\times 10^5$ m·s$^{-1}$). Along the *X*-*S* ($k_y$) and *X*-*U* ($k_z$) directions, the Dirac fermion does not show significant change, as evidenced by the constant energy contour of $E_F$ in the $k_x$ - $k_y$ (Fig. 2**d**) and $k_x$ - $k_z$ plane (Fig. 2**f**), as well as the evolution of the

Dirac fermion dispersion measured at different $k_y$ values (Fig. 2**e**) and different photon energies (Fig. 2**g**). The clear dispersion along $k_x$ direction and dispersionless behavior along $k_y$ and $k_z$ directions demonstrate the 1D Dirac fermion behavior.

**Dimension-crossover of the Dirac fermion.**

Profoundly, we observed the shape of the 1D Dirac fermion shows obvious change with the silicon deficiency $x$ (the $x$ level is confirmed by BZ size measured from ARPES as well as the STM results, see Supplementary Figure 3). As the $x$ value is reduced ($n$ is increased), the metallic $NbTe_2$ chains are aligned closer and start to interact with each other by increasing the overlap between the electron wavefunctions from adjacent chains (e.g. the average distance between $NbTe_2$ chains in samples with $x$ = 0.43 ($n$ = 3)/$x$ = 0.33 ($n$ = 1) are 27.7 Å/12.1 Å, respectively, see Fig. 3**a-b**). Previous STM results report such phenomenon with metallic chain distance varies with different $x$ values[63] (also see Fig. 1**d**). The interchain coupling allows electrons to hop between chains and drives the 1D to 2D dimension-crossover (see schematic in Fig. 3**c-d**). Such dimension-crossover is clearly demonstrated in the constant energy contour near $E_F$ which shows the Dirac cone feature as two parallel lines for $x$ = 0.43 ($n$ = 3) samples (Fig. 3**e**), and alternating shapes (rounded circle to thin line) from the $\bar{S}$ to $\bar{X}$ points for $x$ = 0.40 ($n$ = 2) samples (Fig. 3**h**). The evolution of the Dirac cone dispersion along the $k_y$ direction could be found in Fig. 3**e-j**. As for $x$ = 0.43 ($n$ = 3) sample, the Dirac velocity is 1.07 eV·Å (~1.63× $10^5$ m·s$^{-1}$) along $\bar{\Gamma}-\bar{X}-\bar{\Gamma}$ (Fig. 3**f**) and 0.98 eV·Å (~1.49× $10^5$ m·s$^{-1}$) at $k_y$ = -0.30 Å$^{-1}$ (Fig. 3**g**); while for $x$ = 0.40 ($n$ = 2) sample, the Dirac velocity is 1.44 eV·Å (~2.19× $10^5$ m·s$^{-1}$) along $\bar{\Gamma}-\bar{X}-\bar{\Gamma}$ (Fig. 3**i**) and 0.81 eV·Å (~1.23×

$10^5$ m·s$^{-1}$) at $k_y = 0.30$ Å$^{-1}$ (Fig. 3**j**). The rapid change of Dirac velocity is consistent with the schematic in Fig. 3**c-d** and suggests the 1D to 2D crossover of the Dirac fermions observed.

**Dimension-crossover and the Dirac SSH model of the Dirac fermion.**

The Dirac velocity evolution across the BZ of the two *x* ratio samples are summarized in Fig. 4**a** (Dirac velocity evolution of other *x* ratios is shown in Supplementary Figure 5). Clearly, we could see the Dirac fermion velocity exhibits a periodic variation from $\bar{X}$ to $\bar{S}$ with a difference ($\frac{V_{max} - V_{min}}{V_{max}}$) of ~55.1% for $x = 0.40$ ($n = 2$) sample. Meanwhile, there is much smaller variation (~17.7%) in the Dirac velocity for $x$ =0.43 ($n = 3$) sample, demonstrating the dimension-crossover.

Such behavior could be captured by the following simple model (detailed description could be found in Supplementary Discussion). The two low-energy bands that form the Dirac nodal line are mainly contributed from orbitals at Nb sites in the $NbTe_2$ chains. We first construct a model for a single chain. As illustrated in Fig. 4**b**, the unit cell of the zigzag shaped chain contains two sites *A* and *B* (which may be thought of as corresponding to the Nb sites, although such a correspondence is not necessary). Putting one active orbital per site and denoting the hopping strength between neighboring sites by *t*, the single-chain model takes the form of

$$\mathcal{H}_{\text{Chain}} = \begin{bmatrix} 0 & t(1 + e^{-\mathrm{i}k_\mathrm{x}a}) \\ t(1 + e^{\mathrm{i}k_\mathrm{x}a}) & 0 \end{bmatrix}, \qquad (1)$$

where we assume the chain is orientated along the *x* direction with a lattice constant *a*. This single chain model looks similar to the famous SSH model. However, there is a crucial

distinction: the system here has an important constraint from the glide mirror symmetry $\widetilde{M}_{\mathrm{y}}$. As indicated in Fig. 4**d**, $\widetilde{M}_{\mathrm{y}}$ requires that the sites $A$ and $B$ must be equivalent and the hopping strengths between all neighboring sites must be the same. The algebraic relation $(T\widetilde{M}_{\mathrm{y}})^2 = -1$ for $k_{\mathrm{x}}a = \pi$ further dictates that the bands must form Kramers like degeneracy at the BZ (Fig. 4**e**) boundary. In our single-chain model (1), we have

$$\widetilde{M}_{\mathrm{y}} = \begin{bmatrix} 0 & e^{-\mathrm{i}k_{\mathrm{x}}a} \\ 1 & 0 \end{bmatrix}, \tag{2}$$

and the spectrum $E = \pm 2t\cos\left(\frac{k_{\mathrm{x}}a}{2}\right)$ is gapless with a 1D Dirac type crossing at $k_{\mathrm{x}}a = \pi$ (Fig. 4**f**). This is in contrast with the conventional SSH model, where dimerization is allowed due to the lack of this glide mirror symmetry and it makes the spectrum typically gapped. To highlight the distinction, we call the single-chain model (1) a Dirac SSH model. Next, we extend this Dirac SSH model to 2D by forming an array of the chains. Denoting the lattice period along the *y* direction (inter-chain distance) by *b* and including the inter-chain coupling *t'*, as illustrated in Fig. 4**c**, our 2D Dirac SSH model can be expressed as $\mathcal{H}_{\mathrm{2D}} = \mathcal{H}_{\mathrm{chain}} + \mathcal{H}_{\mathrm{inter}}$ with the inter-chain coupling

$$\mathcal{H}_{\mathrm{inter}} = \begin{bmatrix} 0 & t'e^{-\mathrm{i}k_{y}b}\left(1 + e^{-\mathrm{i}k_{\mathrm{x}}a}\right) \\ t'e^{\mathrm{i}k_{y}b}\left(1 + e^{\mathrm{i}k_{\mathrm{x}}a}\right) & 0 \end{bmatrix}. \tag{3}$$

Importantly, the inter-chain coupling cannot open a gap in the Dirac spectrum. Because $\widetilde{M}_{\mathrm{y}}$ is still preserved, no matter how big *t'* is, the 2D model always has a symmetry enforced Dirac nodal line at the BZ boundary $k_{\mathrm{x}}a = \pi$. Nevertheless, the inter-chain coupling does change the dispersion around the nodal line. The spectrum of our 2D model can be readily found to be

$$E = \pm 2\cos\frac{k_x a}{2}\sqrt{t^2 + 2tt'\cos k_y b + t'^2}. \quad (4)$$

Then the Fermi velocity around the Dirac line in the $k_x$ direction is given by

$$V_f|_{k_x a = \pm\pi} = \pm\frac{a}{\hbar}\sqrt{t^2 + 2tt'\cos k_y b + t'^2}, \quad (5)$$

which depends on $k_y$ when $t'$ is nonzero (Fig. 4**g**). Apparently, the magnitude of the velocity varies along the nodal line, as illustrated in Fig. 4**h**, in an interval $\left[\left||t| - |t'|\right|\frac{a}{\hbar}, \left||t| + |t'|\right|\frac{a}{\hbar}\right]$ (See Supplementary Figure 6-10 for more details of the models). This captures the qualitative features observed in our experiment. The discussion shows that our proposed Dirac SSH model offers a good starting point for understanding the physics in $NbSi_xTe_2$ and for studying general dimension-crossover of Dirac fermions.

## DISCUSSION

In summary, we systematically investigated the electronic structure of the composition-tunable layered compounds $NbSi_xTe_2$ ($x$ = 0.40 and 0.43) and confirmed the 1D Dirac fermion originated from the $NbTe_2$ chains on the surface in the $NbSi_xTe_2$ ($x$ = 0.43) system. Dirac nodal line structure protected by the combined $\widetilde{M_y}$ and time-reversal symmetry $T$ proves the $NbSi_xTe_2$ system as a topological semimetal which is consistent with the ab-initio calculations. The interaction between $NbTe_2$ chains increases as $x$ decreases, and Dirac fermion goes through a dimension-crossover from 1D to 2D which can be explained by a Dirac SSH model. A tunable and unidirectional 1D Dirac electron system is demonstrated in our experiment, which offers a versatile platform for the exploration of intriguing 1D physics and device applications.

# METHODS

## Sample preparation

High-quality $NbSi_xTe_2$ ($x$ = 0.33 to 0.5) single crystals were grown by the chemical vapor transport method by using iodine ($I_2$) as the transport agent. For crystal growth, the polycrystalline powders were synthesized by the direct solid-state reaction using the stoichiometric mixture of high-purity Nb (Alfa Aesar 99.99%), Si (Alfa Aesar 99.999%), and Te powder (Alfa Aesar 99.999%) as staring materials. The ground mixture was sealed in a quartz tube under high vacuum, put in a box furnace, and subsequently heated at 900 °C for several days to prepare the polycrystalline samples. Then the obtained powder and high-purity $I_2$ (Alfa Aesar, 99.9985%) powder were mixed, ground, sealed in another evacuated quartz tube and placed in a double-zone furnace. The temperature profile in the furnace were set to be 850 ℃ (growth side) − 950 ℃ (source side) to grow crystals. After kept this temperature difference for over 10 days, the sheet-like crystals with metallic luster were obtained.

## Angle-resolved photoemission spectroscopy

Synchrotron-based ARPES measurements were performed at the Beamline 10.0.1 of the Advanced Light Source (ALS), Lawrence Berkeley National Laboratory. The samples were cleaved in situ and measurements were performed with temperature of 20 K under UHV below $3 \times 10^{-11}$ Torr. Data were collected by a Scienta R4000 analyzer. The total energy and angle resolutions were 10 meV and 0.2°, respectively.

# DATA AVAILABILITY

The data that support the findings of this study are available from the corresponding authors upon request. Database for material sciences hosted at Institute of Physics and Computer Network Information Center, Chinese Academy of Sciences contains the crystallographic data of this work. These data can be obtained free of charge via http://materiae.iphy.ac.cn/materials/MAT00003867 and http://materiae.iphy.ac.cn/materials/MAT00010455.

## ACKNOWLEDGEMENTS

We thank W. Li, F. Yang and L. Zhang for the helpful discussion. We acknowledge the financial support from the National Key R&D program of China (Grants No.2017YFA0305400), the National Natural Science Foundation of China (No. 51902152) and Singapore MOE AcRF Tier 2 (MOE-T2EP50220-0011). This research used resources of the Advanced Light Source, a U.S. DOE Office of Science User Facility under contract No. DE-AC02-05CH11231.

## COMPETING INTERESTS

The authors declare no competing interests.

## AUTHOR CONTRIBUTIONS

Z.L. and Y.C. supervised the project. J.Z. performed the ARPES measurement with the help of A.L., S.-K.M. and C.C.. S.A.Y. and X.F. provided the theoretical modeling support. Y.L. and Y.C. synthesized the crystals. All authors discussed the results and contributed to the paper.

## ADDITIONAL INFORMATION

**Supplementary information** is available for this paper at

**Correspondence** and requests for materials should be addressed to Y.C. or Z.L.

## Figure legends

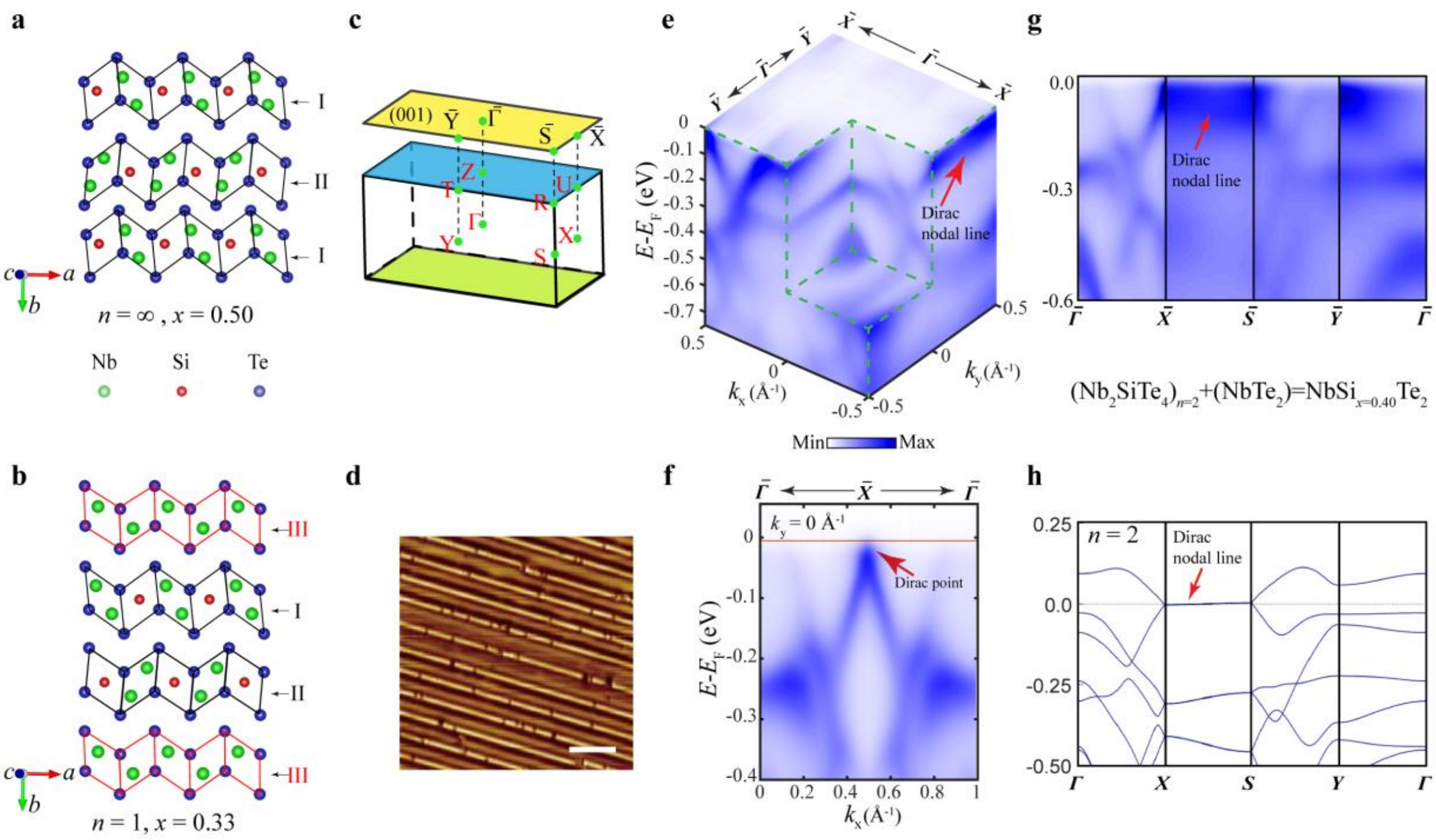

**Fig. 1: Crystal structure and characterizations of $NbSi_xTe_2$. a** Crystal structure of $Nb_2SiTe_4$ ($n = \infty$, $x = 0.5$). Type I and II are two kinds of chains along the *a*-axis as building blocks. **b** Crystal structure of $Nb_3SiTe_6$ ($n = 1$, $x = 0.33$). Type I, II and III are three kinds of chains as building blocks. **c** The Brillouin zone (BZ) and its projection to the (001) surface of $Nb_3SiTe_6$. **d** STM topography image of $NbSi_xTe_2$. The bright lines show the metallic $NbTe_2$ chain (type III). Scale bar represents 10nm. **e** 3D volume plot of the overall band structure of $x = 0.40$ ($n = 2$) sample with Dirac nodal line labelled. **f** The photoemission spectrum along the high symmetry $\bar{\Gamma} - \bar{X} - \bar{\Gamma}$ directions of $x = 0.40$ ($n = 2$) sample. Dirac point (DP) is labelled on the spectrum. **g** The photoemission spectrum along the high symmetry $\bar{\Gamma} - \bar{X} - \bar{S} - \bar{Y} - \bar{\Gamma}$ direction of $x = 0.40$ ($n = 2$) sample. Dirac nodal line is labelled. **h** The ab-initio calculated band structure of $x = 0.40$ ($n = 2$) sample from Ref. **63.** Dirac nodal line is labelled.

Data in Fig. 1**e-g** were collected at *hv* = 56 eV with sample temperature 20 K. For STM topography measurement, the bias voltage is 0.7 V, and the tunneling current is 50 pA. The sample temperature is 4.2 K. Data from Ref. **58** (Reprinted (adapted) with permission from {Wang, B. *et al.* One-dimensional metal embedded in two-dimensional semiconductor in $Nb_2Si_{x-1}Te_4$. *ACS Nano* **15**, 7149-7154 (2021).}. Copyright {2022} American Chemical Society)**.**

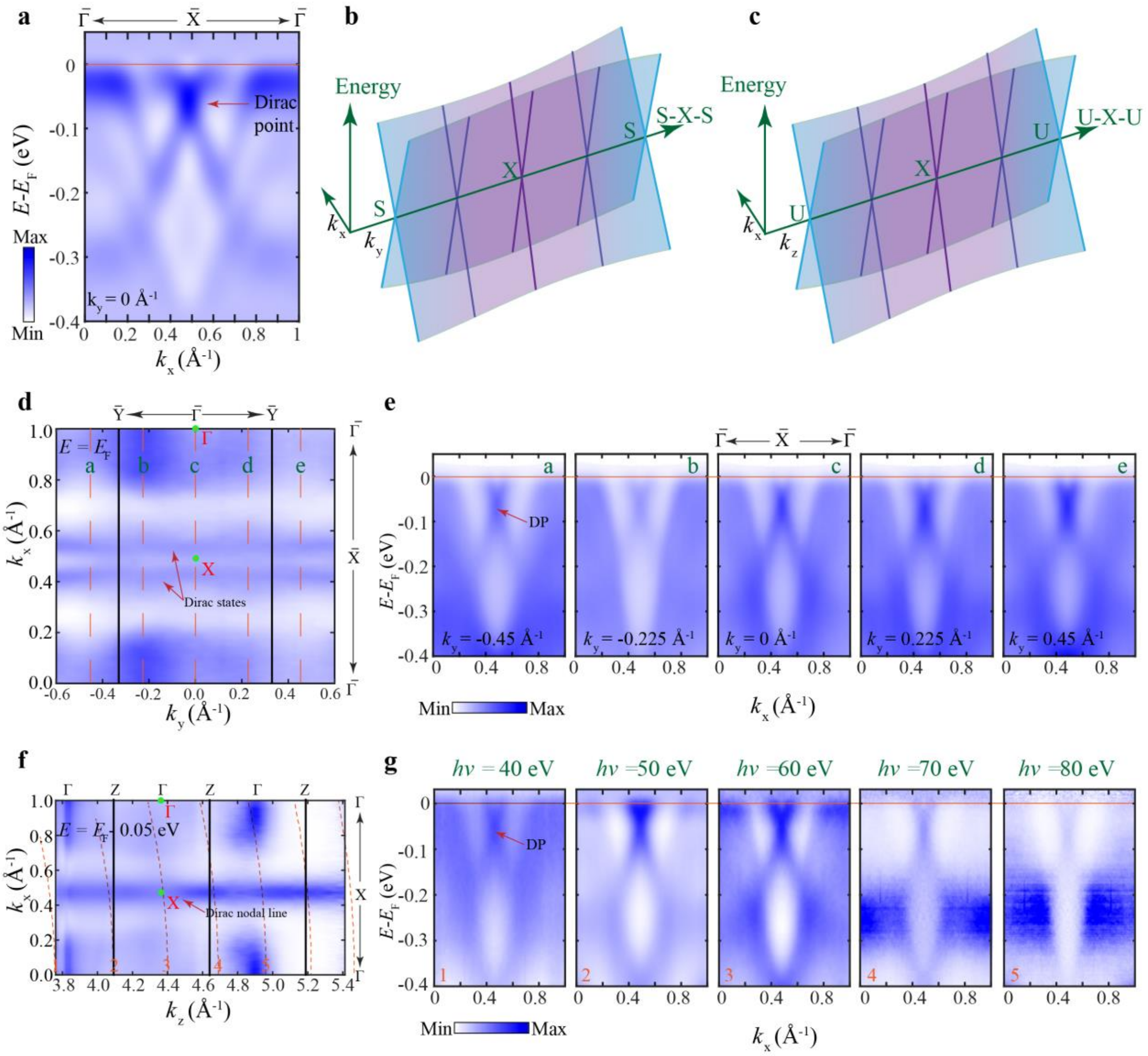

**Fig. 2: Band structure of the 1D Dirac fermion of $x$ = 0.43 ($n$ = 3) sample. a** Second derivative of the photoemission spectrum $\frac{d^2A}{d\omega^2}$ along $\bar{\Gamma} - \bar{X} - \bar{\Gamma}$ with DP labelled. **b-c** Schematic of the 1D Dirac fermion along the $S$-$X$-$S$ ($k_y$) and $U$-$X$-$U$ ($k_z$) directions. **d** The constant energy contour at Fermi energy ($E_F$) in the $k_x - k_y$ plane with Dirac surface sates labelled and theoretical BZ appended (black solid lines). **e** Sliced cut in Fig. 2**d** at different $k_y$ (orange dashed lines indicated in 2**d**) with DP labelled. **f** The energy contour near Fermi energy ($E_F$) of the $k_x - k_z$ plane with Dirac nodal line labelled. **g** Sliced cuts in Fig. 2**f** at different $hv$ ($k_z$ - orange dashed lines) with DP labelled. To make a clear comparison, we symmetrized the dispersion according to the crystal symmetry. The data in Fig. 2**a**, 2**d** were collected at $hv$ = 55 eV.

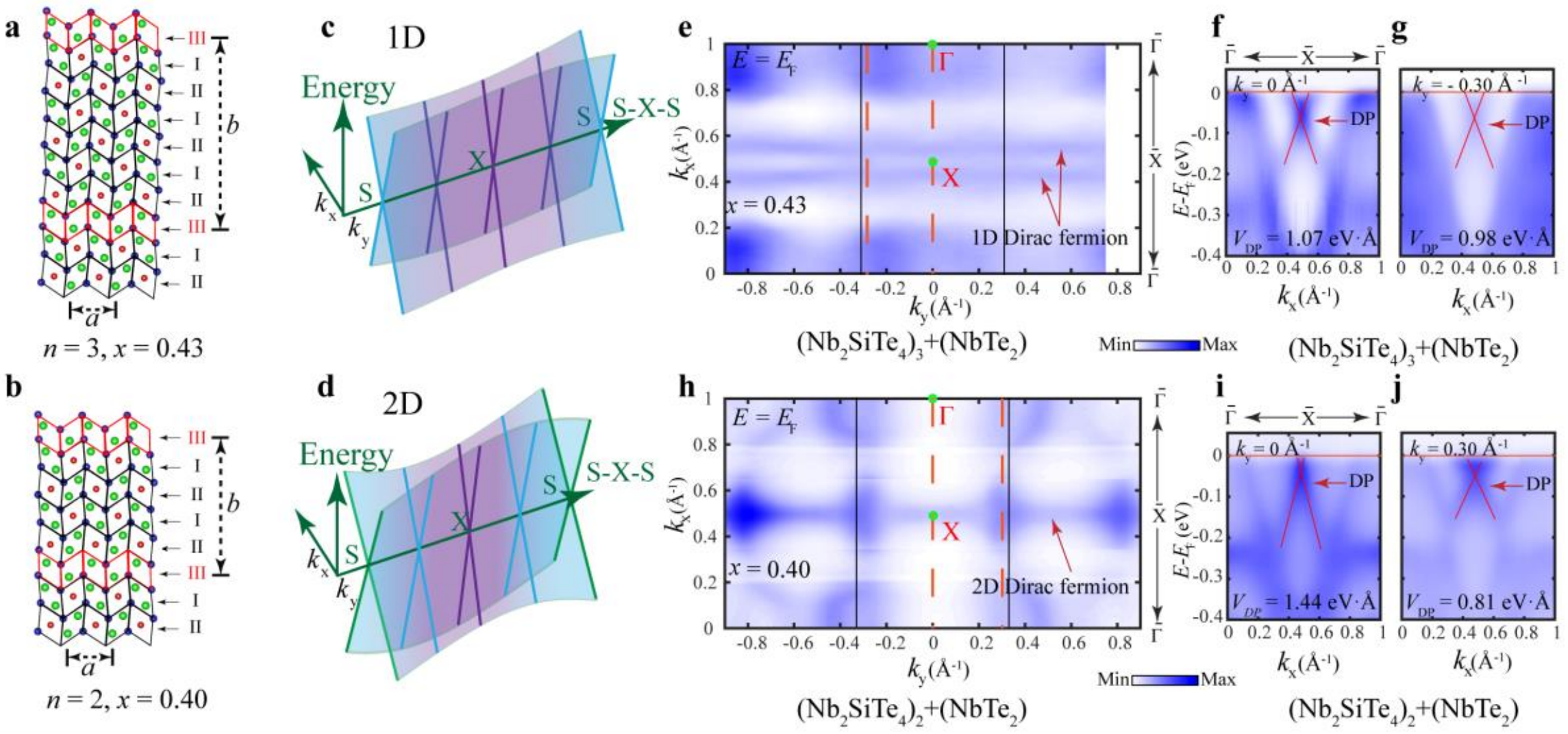

**Fig. 3: Dimension-crossover of the Dirac fermion. a-b** Crystal structure of (**a**) $x$ = 0.43 ($n$ = 3) and (**b**) $x$ = 0.40 ($n$ = 2) sample. **c-d** Schematic of the (**c**) 1D and (**d**) 2D Dirac fermion along the $S$-$X$-$S$ direction ($k_y$). **e** The energy contour near Fermi energy ($E_F$) of the $k_x - k_y$ plane with theoretical BZ appended (black solid lines). **f-g** The spectrum (**f**) along $\bar{\Gamma} -$

$\bar{X} - \bar{\Gamma}$ and the spectrum (**g**) at $k_y$ = -0.30 Å$^{-1}$ (orange dashed lines) with linear fitting result of the Dirac fermion appended for $x$ = 0.43 sample. **h-j** Same as **e-g**, but for $x$ = 0.40 sample. Data in Fig. 3**e-j** were collected at $h\nu$ = 56 eV.

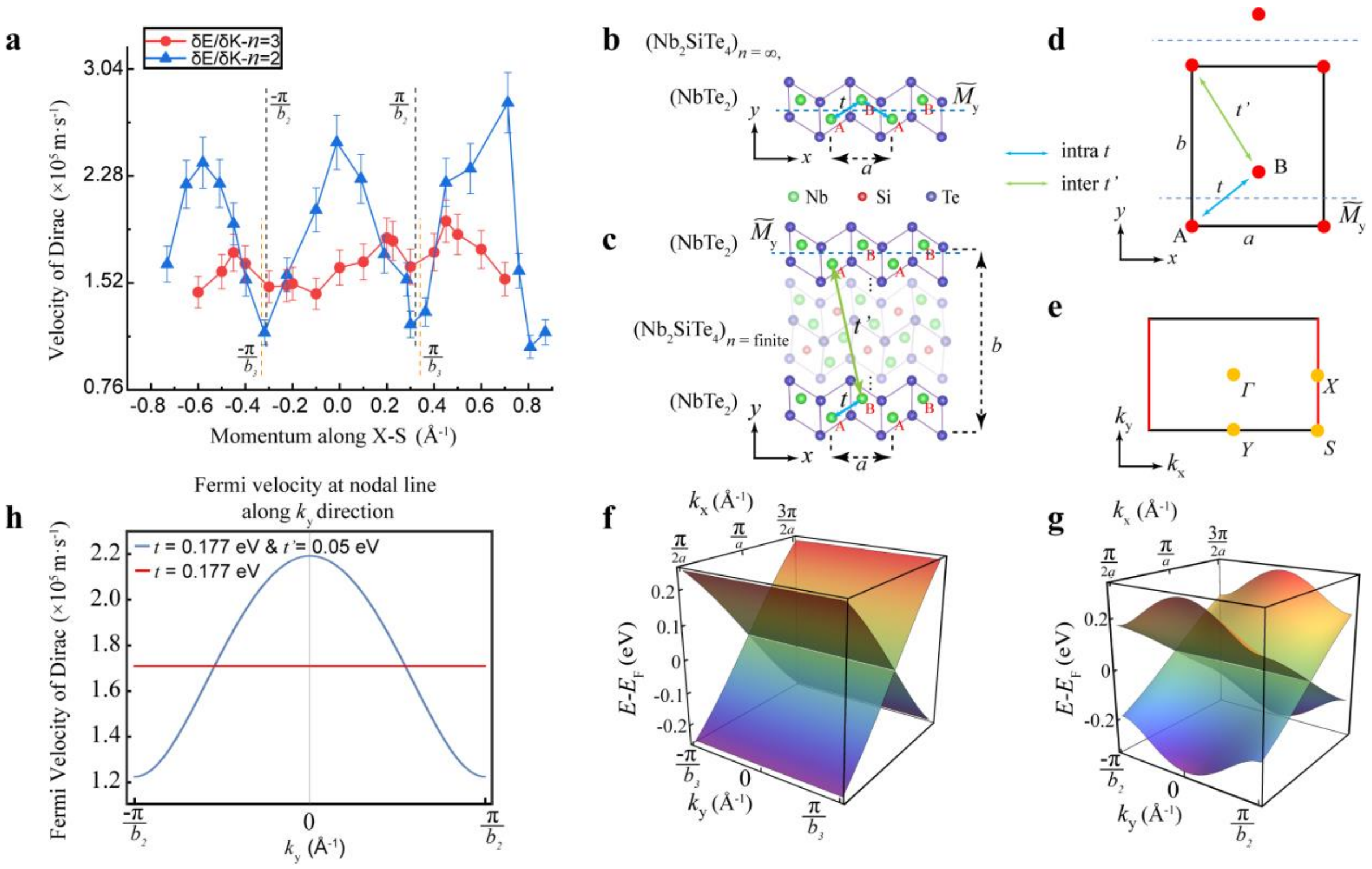

**Fig. 4: Dimension-crossover and the Dirac SSH model of the Dirac fermion. a** The extracted Dirac fermion velocity along $X$-$S$ in $NbSi_xTe_2$ for $x$ = 0.43 ($n$=3) and $x$ = 0.40 ($n$=2) samples with the theoretical BZ period appended. **b** Schematic of a single $NbTe_2$ chain. Sites $A$, $B$ and intra-chain coupling $t$ as well as glide mirror $\widetilde{M_y}$ are labelled. **c** Schematic of the $NbTe_2$ chains array extracted from the $NbSi_xTe_2$ surface. $NbTe_2$ chain distance along the $y$ direction is marked by $b$, intra-chain coupling ($t$) and inter-chain coupling ($t'$) are labelled. **d** Plot of the extracted Dirac SSH model from Fig. 4**c**. Two sites $A$ and $B$ are included in the cell, related by glide mirror $\widetilde{M_y}$. The lattice constants, the intra- and interchain hopping $t$ and $t'$ are labelled. The dash line denotes the glide plane. **e** Corresponding BZ for the Dirac SSH

model. The orange points denote the high symmetry points and red lines denote the location of nodal line at $k_x{\cdot}a = \pm\pi$. **f** 1D Dirac SSH model derived dispersion with $t = 0.177$ eV. **g** 2D Dirac SSH model derived dispersion with $t = 0.177$ eV and $t' = 0.05$ eV. **h** Calculated Fermi velocity of the Dirac fermion along the $k_x$ direction and its evolution along the $k_y$ direction with $t = 0.177$ eV (red curve) and $t = 0.177$ eV, $t' = 0.05$ eV (blue curve).